\documentclass[3p,times]{elsarticle}

\usepackage{ecrc}

\usepackage{amsmath}
\usepackage{amssymb}

\volume{00}

\firstpage{1}

\journalname{Nuclear Physics A}

\runauth{}

\jid{npa}

\jnltitlelogo{Nuclear Physics A}

\usepackage{amssymb}

\biboptions{square,comma,numbers,sort&compress}

\usepackage[figuresright]{rotating}

\newcommand{\bc}{\begin{center}}
\newcommand{\ec}{\end{center}} 
\newcommand{\be}{\begin{equation}}
\newcommand{\ee}{\end{equation}}
\newcommand{\beq}[1]{\begin{equation} \label{#1}}
\newcommand{\eeq}[1]{\label{#1} \end{equation}}
\newcommand{\bea}{\begin{eqnarray}}
\newcommand{\eea}{\end{eqnarray}}
\newcommand{\beqar}{\begin{eqnarray}} 
\newcommand{\eeqar}[1]{\label{#1} \end{eqnarray}} 
\newcommand{\bit}{\begin{itemize}}
\newcommand{\eit}{\end{itemize}}
\newcommand{\bsp}{\begin{split}}
\newcommand{\esp}{\end{split}}

\newcommand{\lp}{\left(}
\newcommand{\rp}{\right)}

\newcommand{\bk}{\mathbf{k}_\perp}

\newcommand{\bq}{\mathbf{q}_\perp}
\newcommand{\bx}{\mathbf{x}}

\newcommand{\bz}{\mathbf{z}}
\newcommand{\vz}{\mathbf{z}}

\newcommand{\raa}{$ R_{AA} $}
\newcommand{\vtwo}{$ v_{2} $}
\newcommand{\chisq}{$ \chi^2/d.o.f. $}
\newcommand{\amax}{$ \alpha_{max} $}

\begin{document}

\begin{frontmatter}



\dochead{}

\title{The tricky azimuthal dependence of jet quenching at RHIC and LHC via CUJET2.0}


\author[CU]{Jiechen Xu}
\author[CU,LBL]{Alessandro Buzzatti}
\author[CU,LBL,HU]{Miklos Gyulassy}

\address[CU]{Department of Physics, Columbia University, 538 West 120th Street, New York, NY, USA}
\address[LBL]{Nuclear Science Division, Lawrence Berkeley National Laboratory, 1 Cyclotron Road, Berkeley, CA, USA}
\address[HU]{Institute for Particle and Nuclear Physics, Wigner RCP, HAS,
1121 Budapest, XII. Konkoly Thege Mikl\'{o}s \'{u}t 29, Hungary}

\begin{abstract}
High transverse momentum neutral pion and charged hadron suppression pattern with respect to reaction plane at RHIC and LHC energies in central and semi-peripheral AA collisions are studied in a perturbative QCD based model, CUJET2.0. CUJET2.0 has dynamical DGLV radiation kernel and Thoma-Gyulassy elastic energy loss, with both being generalized to including multi-scale running strong coupling as well as energy loss probability fluctuations, and the full jet path integration is performed in a low $p_T$ flow data constrained medium which has 2+1D viscous hydrodynamical expanding profile. We find that in CUJET2.0, with only one control parameter, $\alpha_{max}$, the maximum coupling strength, fixed to be 0.26, the computed nuclear modification factor $R_{AA}$ in central and semi-peripheral AA collisions are consistent with RHIC and LHC data at average $\chi^2/d.o.f.<1.5$ level. Simultaneous agreements with high $p_T$ azimuthal anisotropy $v_2$ data are acquired given average $\alpha_{max}$ over in-plane and out-of-plane paths varying as little as 10\%, suggesting a non-trivial dependence of the high $p_T$ single particle $v_2$ on the azimuthally varied strong coupling.
\end{abstract}

\begin{keyword}
Heavy ion phenomenology \sep Jets


\end{keyword}

\end{frontmatter}


\section{Introduction}
\label{intro}

Quantitatively predicting jet quenching observables in relativistic heavy-ion collisions at RHIC and LHC is an important concentration of perturbative QCD (pQCD) based jet energy loss models. Given the complexity of jet medium dynamics and bulk flow fields, clarifying the underlying parton medium interaction mechanism and the evolution profile of quark-gluon plasmas (QGP) is a crucial. Recent jet quenching data from RHIC \cite{PHENIX,STAR} and LHC \cite{ALICE,ATLAS,CMS}, in particular, neutral pion $(\pi^0)$ and charged hadron $(h^\pm)$ nuclear modification factor \raa~and single particle azimuthal anisotropy \vtwo, have provided key constraints for pQCD jet tomography models.  

Within the pQCD framework, AMY, ASW and HT models well-explained the recent experimentally observed leading hadron \raa~at both RHIC and LHC, however significantly underestimated high transverse momentum ($p_T$) elliptic flow \vtwo~\cite{Barbara}.
To gain simultaneous consistency with $R^{h}_{AA}(p_T>5{\rm GeV},|\eta|<1;\sqrt{s},b)$ and $v^{h}_{2}(p_T>5{\rm GeV},|\eta|<1;\sqrt{s},b)$ data in various $\sqrt{s}$, $b$ setup of AA collisions with minimally introduced tuning degrees of freedom is therefore one of the most imperative phenomenological goals for pQCD hard probe models.

The CUJET2.0 \cite{CUJET2.0} model is developed in the pQCD scenario based on the (D)GLV opacity expansion theory \cite{GLV,GLV+}. It features: (1) numerical evaluation of DGLV opacity series for radiative jet energy loss in dynamical QCD medium; (2) inclusion of fluctuating Thoma-Gyulassy (TG) elastic energy loss; (3) coupling radiation and scattering kernels to state-of-the-art 2+1D viscous hydrodynamical fluid fields; (4) inclusion of multi-scale running strong coupling in both inelastic and elastic energy loss sectors; (5) full jet path integration with realistic geometry fluctuations; (6) convolution over numerical pQCD initial production spectra of all flavors; and (7) convolution over jet fragmentation functions and evaluation of heavy flavor lepton spectra.

The kernel of CUJET2.0 energy loss model for induced gluon bremsstrahlung is the running coupling DGLV opacity series. To the first order in opacity, it has the form:
\be
\begin{split}
x_E \frac{dN_g^{n=1}}{dx_E}(\bx_0,\phi) = &\; \frac{18 C_R}{\pi^2} \frac{4+n_f}{16+9n_f} \int{d\tau}\; \rho(\bz) \int{d^2\bk} \int{d^2\bq}\;{\frac{\alpha_s^2(\bq^2)(f_E^2-f_M^2)}{(\bq^2+f_E^2\mu^2(\vz))(\bq^2+f_M^2\mu^2(\vz))}}\\
&\times\;\alpha_s ( \frac{\bk^2}{x_+ (1-x_+)} )\;{\frac{-2(\bk-\bq)}{(\bk-\bq)^2+\chi^2(\bz)} \left[ \frac{\bk}{\bk^2+\chi^2(\bz)} - \frac{(\bk-\bq)}{(\bk-\bq)^2+\chi^2(\bz)} \right] }\\
&\times\;{\left[ 1-\cos\lp\frac{(\bk-\bq)^2+\chi^2(\bz)}{2 x_+ E } \tau\rp\right]}\;{\lp \frac{x_E}{x_+} \rp J(x_+(x_E))}
\; \; ,
\label{rcCUJETDGLV}
\end{split}
\ee
where $ \vz=\lp x_0+\tau\cos\phi,y_0+\tau\sin\phi; \tau\rp$ is the coordinate of the jet in the transverse plane; $\rho(\vz)$ and $T(\vz)$ are the local number density and temperature of the medium; $\chi^2(\vz)=M^2 x_+^2+m_g^2(\vz)(1-x_+)$, and squared gluon plasmon mass $ m_g^2(\vz)=f_E^2 \mu^2(\vz) / 2 $, 1-HTL Debye mass $ \mu(\vz) = g(\vz)T(\vz)\sqrt{1+n_f/6} $, and $ g(\vz)=\sqrt{4\pi\alpha\lp4T^2(\vz)\rp} $. The running strong coupling $\alpha_s(Q^2)$ has Zakharov's 1-loop pQCD running that is cutoff in the infrared when coupling strength reaches a maximum value $\alpha_{max}$: $\alpha_s\rightarrow\alpha_s(Q^2) = \alpha_{max}\;(Q^2 \le Q^2_{min}),\;{4\pi}[\beta_0\log(Q^2/\Lambda^2_{QCD})]^{-1}\;(Q^2 > Q^2_{min})$ \cite{Zakharov}. 
The minimum running scale $Q^2_{min}$ is fixed by $Q^2_{min}=\Lambda^2_{QCD}\exp\left\lbrace {4\pi}/{9\alpha_{max}}\right\rbrace$ according to $\alpha_{max}$. The assumed upper bound of strong coupling strength \amax, along with HTL deformation parameters $f_E$ and $f_M$ in the effective scattering potential, span the parameter space of CUJET2.0. In the elastic sector, TG formula is used:
\be\label{rcCUJETElastic}
\frac{dE(\bz)}{d\tau}= - C_R \pi\;\alpha_s(\mu(\bz)^2)\;\alpha_s( E(\bz) T(\bz))\;T(\bz)^2 \lp 1+\frac{n_f}{6} \rp \; \log \left\lbrace  \frac{4T(\bz)\sqrt{E(\bz)^2-M^2}}{\left[ E(\bz)-\sqrt{E(\bz)^2-M^2}+4T(\bz)\right]\mu(\bz)} \right\rbrace\;.
\ee
A Poisson ansatz is assumed in the radiative sector for incoherent multiple gluon emissions, while a Gaussian fluctuation is applied in the elastic. Total energy loss probability distribution is the convolution of radiative and elastic, it is then convoluted with firstly pQCD pp jet production spectra, secondly Glauber AA initial jet distribution, and finally jet fragmentation functions to get the hadron spectra in AA collisions.

\section{Results and discussions}
\label{result}

The CUJET2.0 model can be coupled to a wide range of bulk evolution profiles. At present stage, the energy loss kernel is integrated with fluid fields generated from VISH \cite{VISH} 2+1D viscous hydro which is constrained by low $p_T$ flow data. We study central b=2.4fm/0-10\% centrality and semi-peripheral b=7.5fm/10-30\% centrality AA collisions at RHIC and LHC energies. The HTL deformation parameters $(f_E,f_M)$ in Eq.~\eqref{rcCUJETDGLV} are fixed to be $(1,0)$ to maintain the HTL scenario in a dynamical QCD medium, it makes \amax~the only adjustable parameter in our model.

\begin{figure}[!h]
\bc
\includegraphics[width=0.3\textwidth]{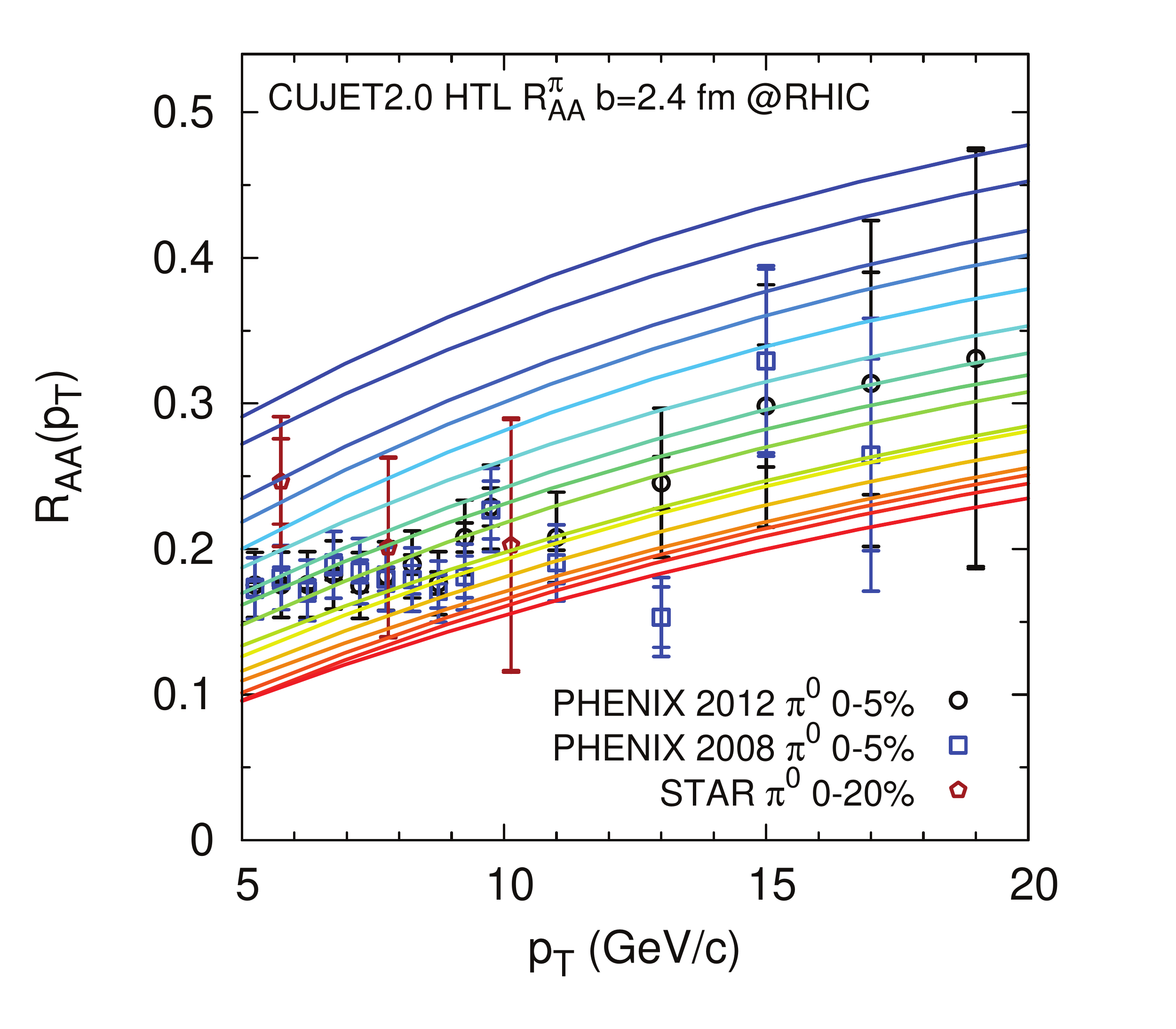}
\includegraphics[width=0.3\textwidth]{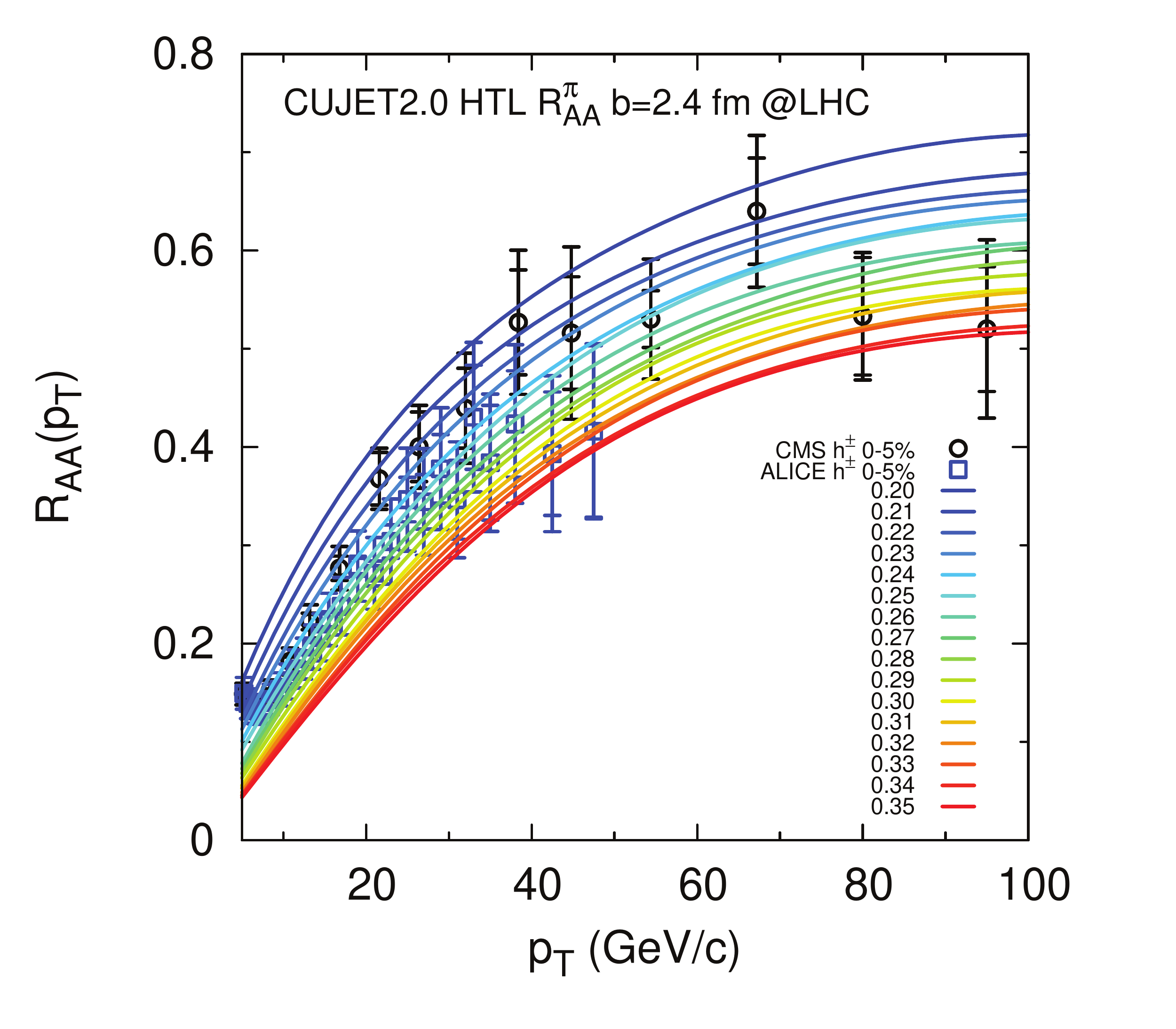}
\includegraphics[width=0.3\textwidth]{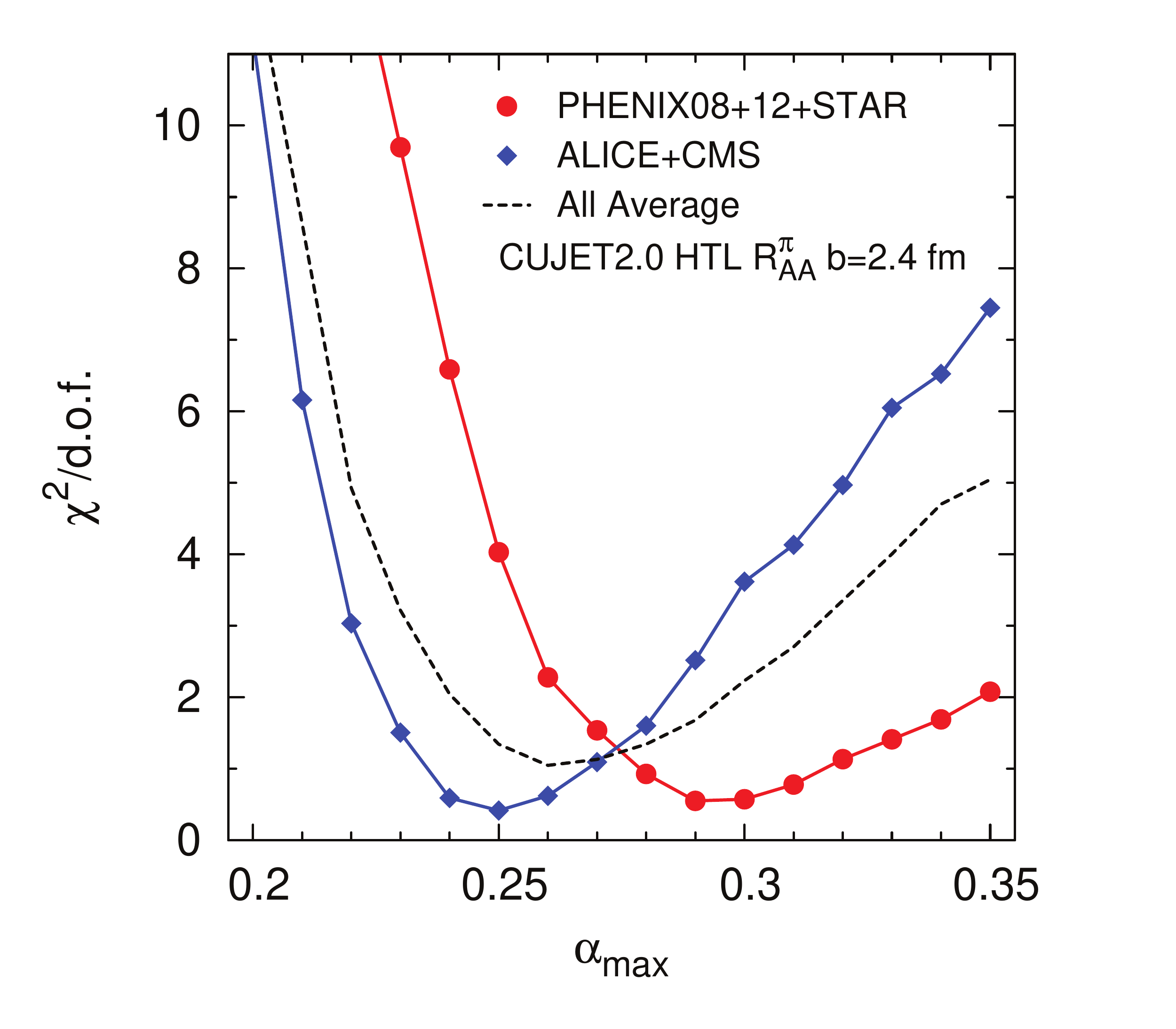}
\vspace{-0.18 in}
\includegraphics[width=0.3\textwidth]{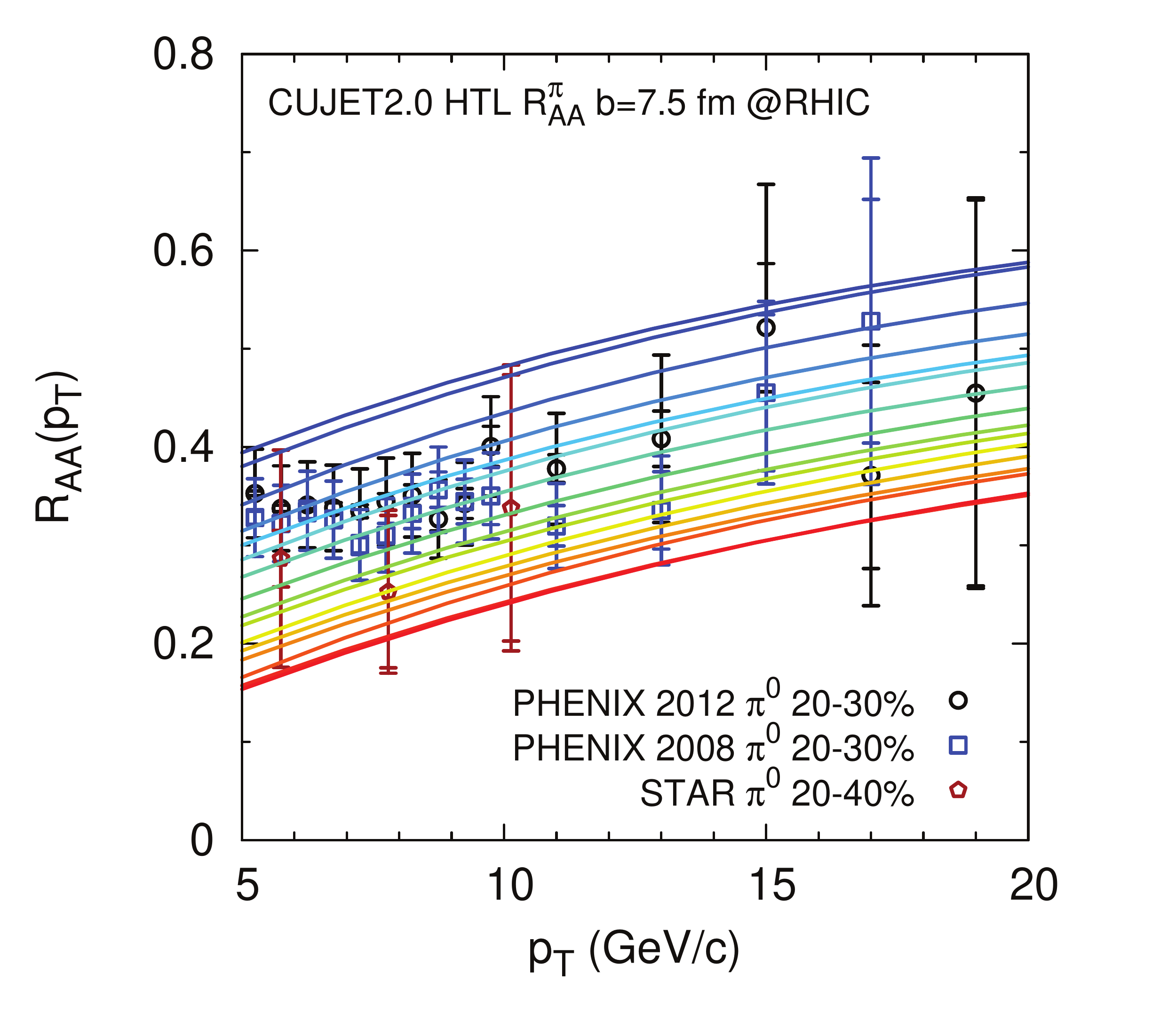}
\includegraphics[width=0.3\textwidth]{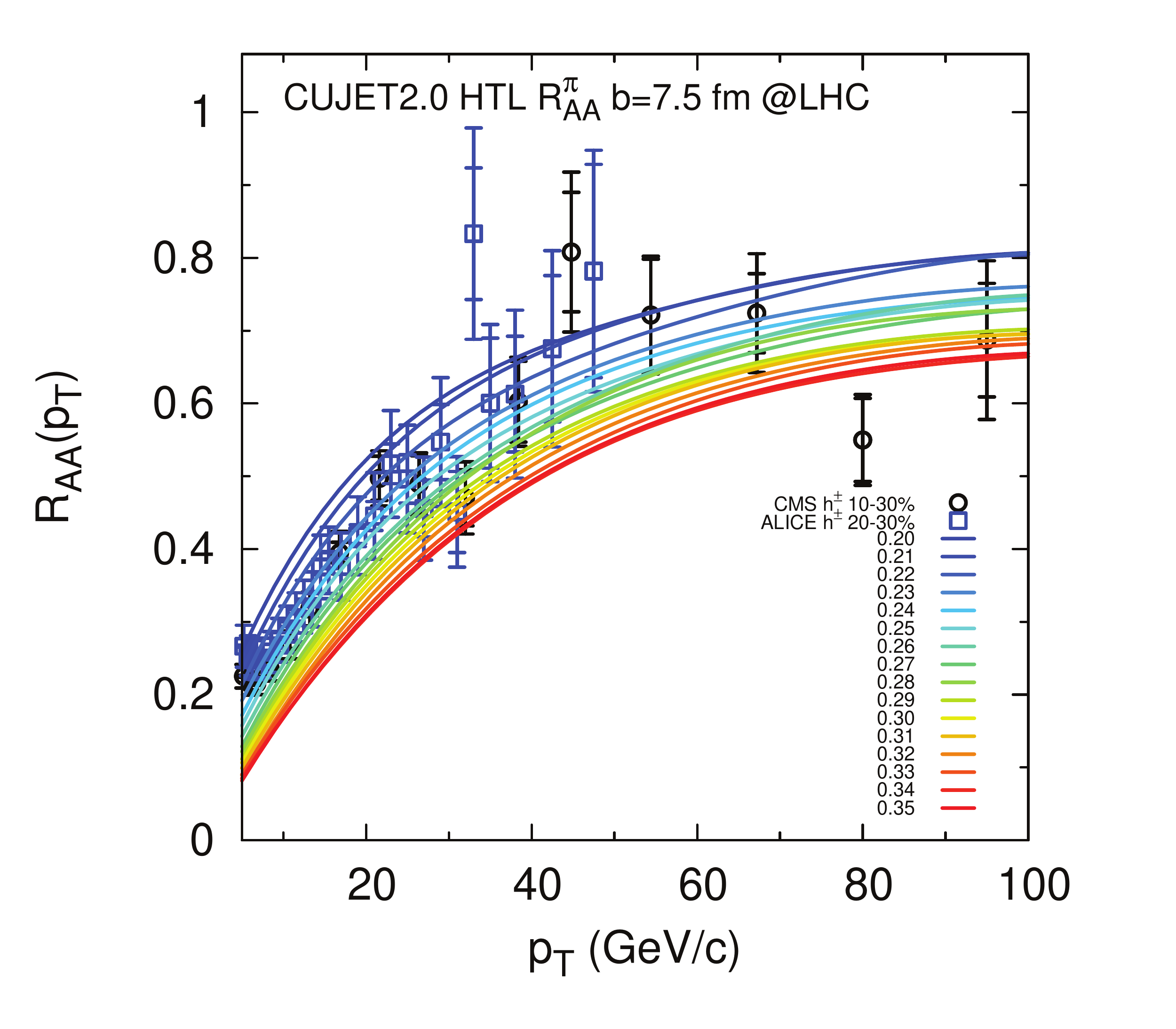}
\includegraphics[width=0.3\textwidth]{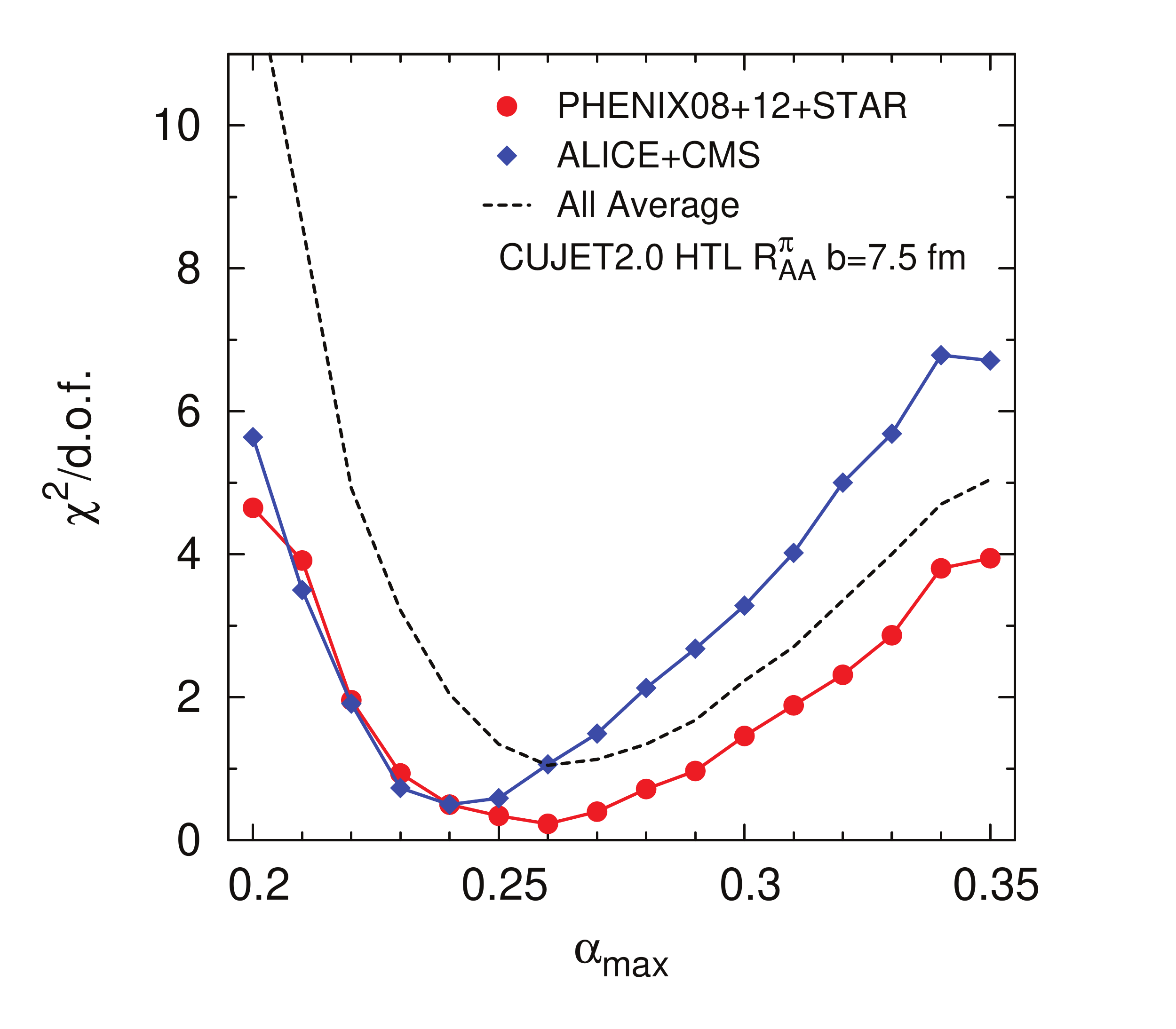}
\caption{
		$ R^\pi_{AA} $ versus $ p_T $ in CUJET2.0 and $\chi^2$ analysis. $ \alpha_{max}$ is increased from $0.20$ to $0.35$ in $0.01$ steps in the dynamical HTL scenario $(f_E=1,f_M=0)$. Azimuthally averaged $ R^\pi_{AA} $ in RHIC Au+Au $\sqrt{s_{NN}}=200 \mbox{GeV}$ b=2.4fm (top left), 7.5fm (bottom left); LHC Pb+Pb $\sqrt{s_{NN}}=2.76 \mbox{TeV}$ b=2.4fm (top middle), 7.5fm (bottom middle) collisions are calculated. The results are compared to measurements of: PHENIX 2008 and 2012 ${\pi^0} R_{AA}(p_T) $ \cite{PHENIX}, STAR ${\pi^0} R_{AA}(p_T) $ \cite{STAR} in Au+Au 200AGeV collisions; ALICE ${h^\pm} R_{AA}(p_T) $ \cite{ALICE}, CMS ${h^\pm} R_{AA}(p_T) $ \cite{CMS} in Pb+Pb 2.76ATeV collisions. VISH2+1 \cite{VISH} bulk evolution profile is used, which has MC-Glauber initial condition, $\tau_0=0.6$ fm/c, s95p-PCE EOS, $\eta/s=0.08$ and $T_f=120$ MeV. In real-time computations, smooth profiles from VISH2+1 are embedded. The path integration in Eq.~\eqref{rcCUJETDGLV} is cutoff at $T({\bf x}_0,\phi,\tau)|_{\tau_{max}}=120$ MeV. $ \chi^2/d.o.f. $ versus $ \alpha_{max} $ is calculated in b=2.4fm (top right), 7.5fm (bottom right) for RHIC (red) and LHC (blue) respectively, and the average over all four \chisq's is plotted as a reference in both panels (dashed black). In terms of average \raa, CUJET2.0 HTL scenario has $ \alpha_{max}=0.25-0.27 $ at average $ \chi^2/d.o.f.<1.5 $ level, and $ \alpha_{max}=0.23-0.30 $ at average $ \chi^2/d.o.f.<2 $ level. The small value of the strong coupling constant is partially resulted from the dominating longer jet path length in the transverse expanding medium, which overrides reduced medium density and induces overall enhanced quenching. (For interpretation of the references to color in this figure legend, the reader is referred to the web version of this article.)
		}
\label{fig:Multi_Alf_HTL}
\ec
\end{figure}

The average \raa~results for RHIC Au+Au $\sqrt{s_{NN}}=200\rm GeV$ and LHC Pb+Pb $\sqrt{s_{NN}}=2.76\rm TeV$ are shown in Fig.~\ref{fig:Multi_Alf_HTL}.
The $ \alpha_{max}$ value rises from $0.20$ to $0.35$ with $0.01$ increment and corresponding $\chi^2$ analysis is performed. With the combination of running coupling and transverse expanding medium effect, the steep rising and subsequent flattening signature of $p_T$ dependent on \raa~at LHC is manifest within the CUJET2.0 framework, as opposed to fixed coupling CUJET1.0. According to the $\chi^2$ analysis, CUJET2.0 HTL $ \alpha_{max}=0.25-0.27 $ model agrees with measurements of $\pi^0/h^\pm\;R_{AA}$ at RHIC and LHC both central and semi-peripheral AA collisions {\it simultaneously} at the level of $ \chi^2/d.o.f.<1.5 $. Using this \amax~range, the effective jet transport coefficient $\hat{q}$ calculated in CUJET2.0 are consistent with $\hat{q}$ extracted from HT-BW, HT-M, MARTINI and McGill-AMY models \cite{qhat}. And if one allows for $ \chi^2/d.o.f.<2 $, $ \alpha_{max}=0.23-0.30 $. The \amax~value in CUJET2.0 is significantly smaller than the case of running coupling CUJET with transverse static Glauber + longitudinal Bjorken expansion, whose $\alpha_{max}=0.4$, suggesting longer jet path length in a transverse expanding medium overrides the diminished density and results in overall more energy loss.

\begin{figure*}[!h]
\bc
\includegraphics[width=0.3\textwidth]{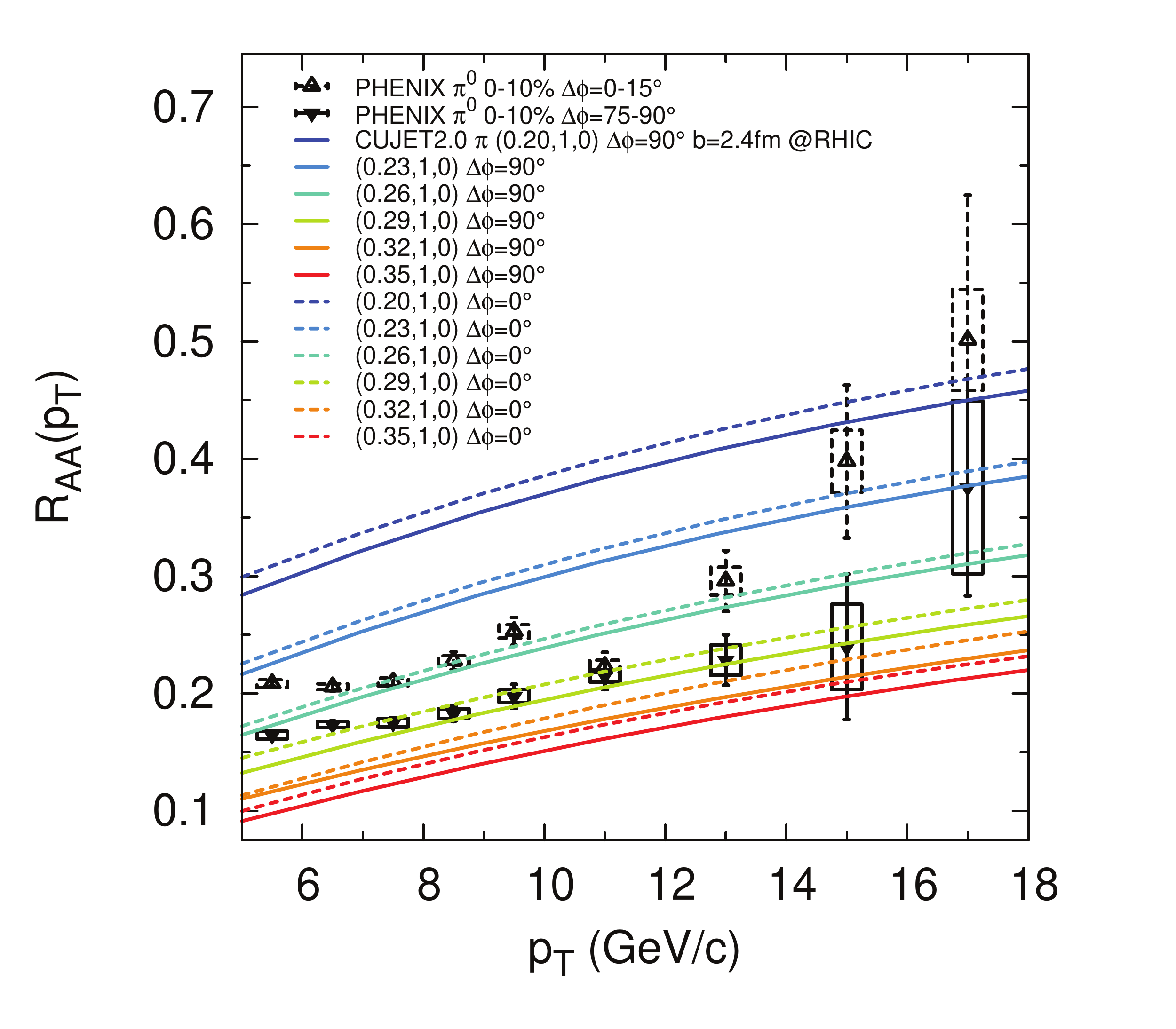}
\includegraphics[width=0.3\textwidth]{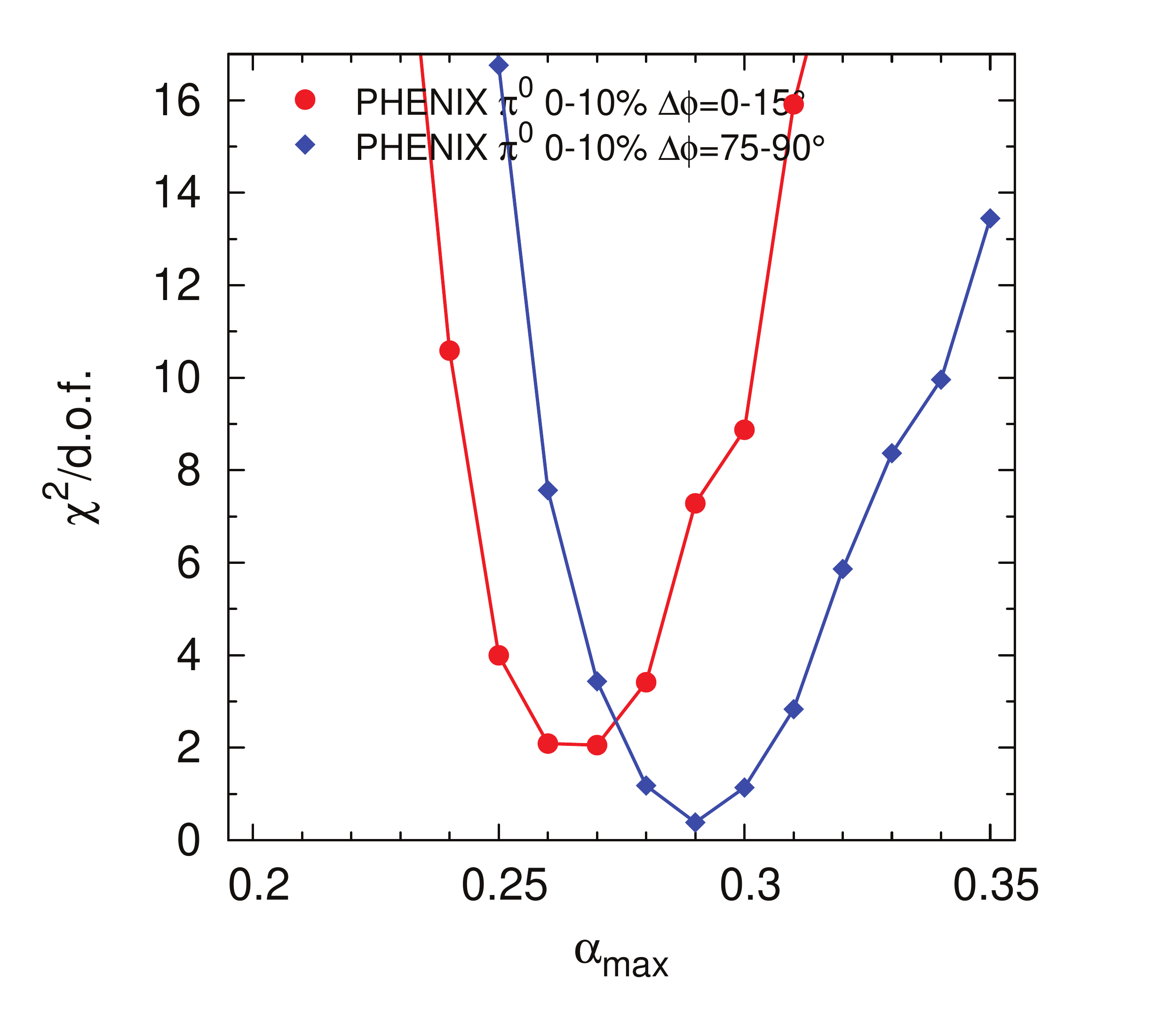}
\includegraphics[width=0.3\textwidth]{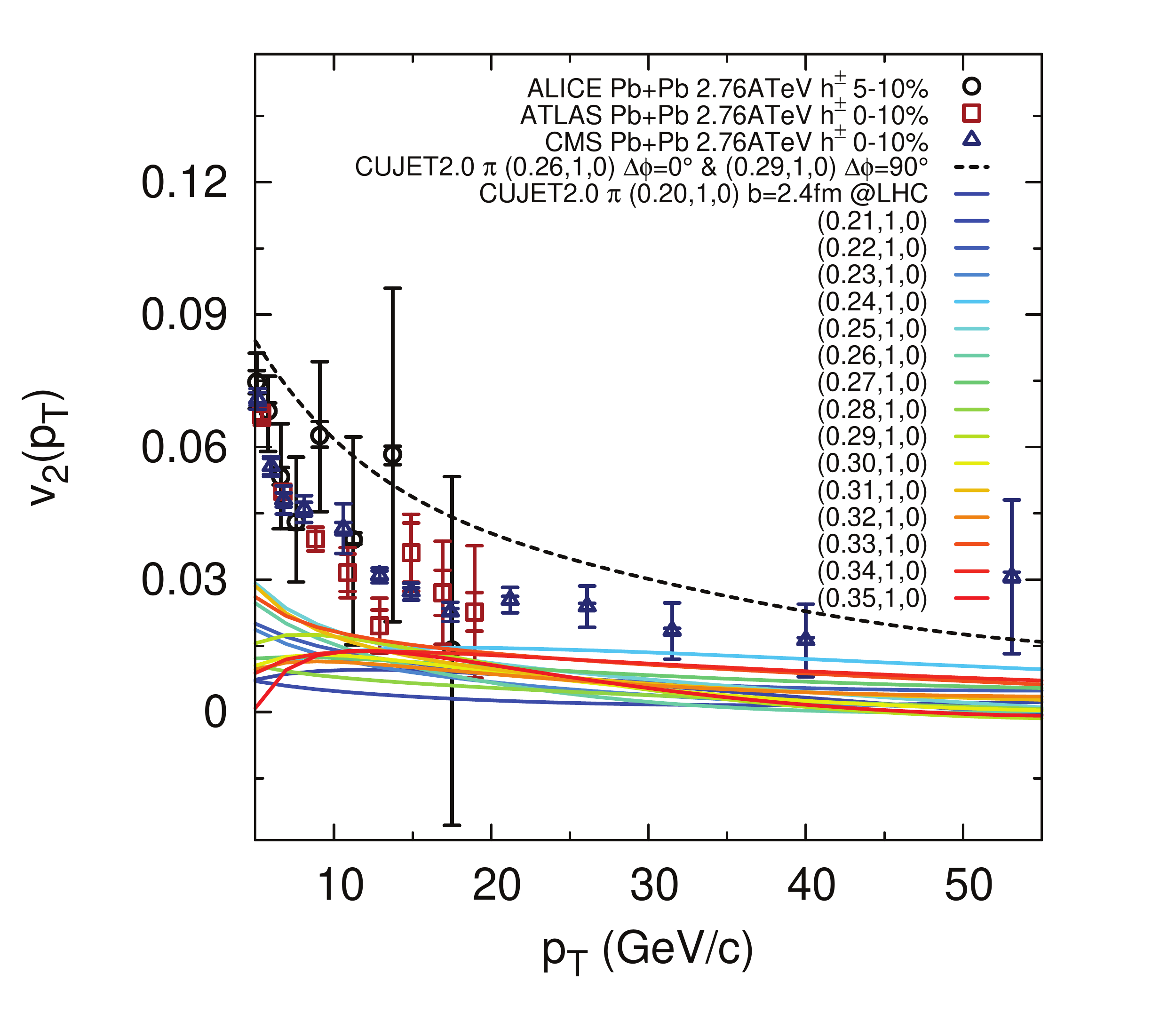}
\includegraphics[width=0.3\textwidth]{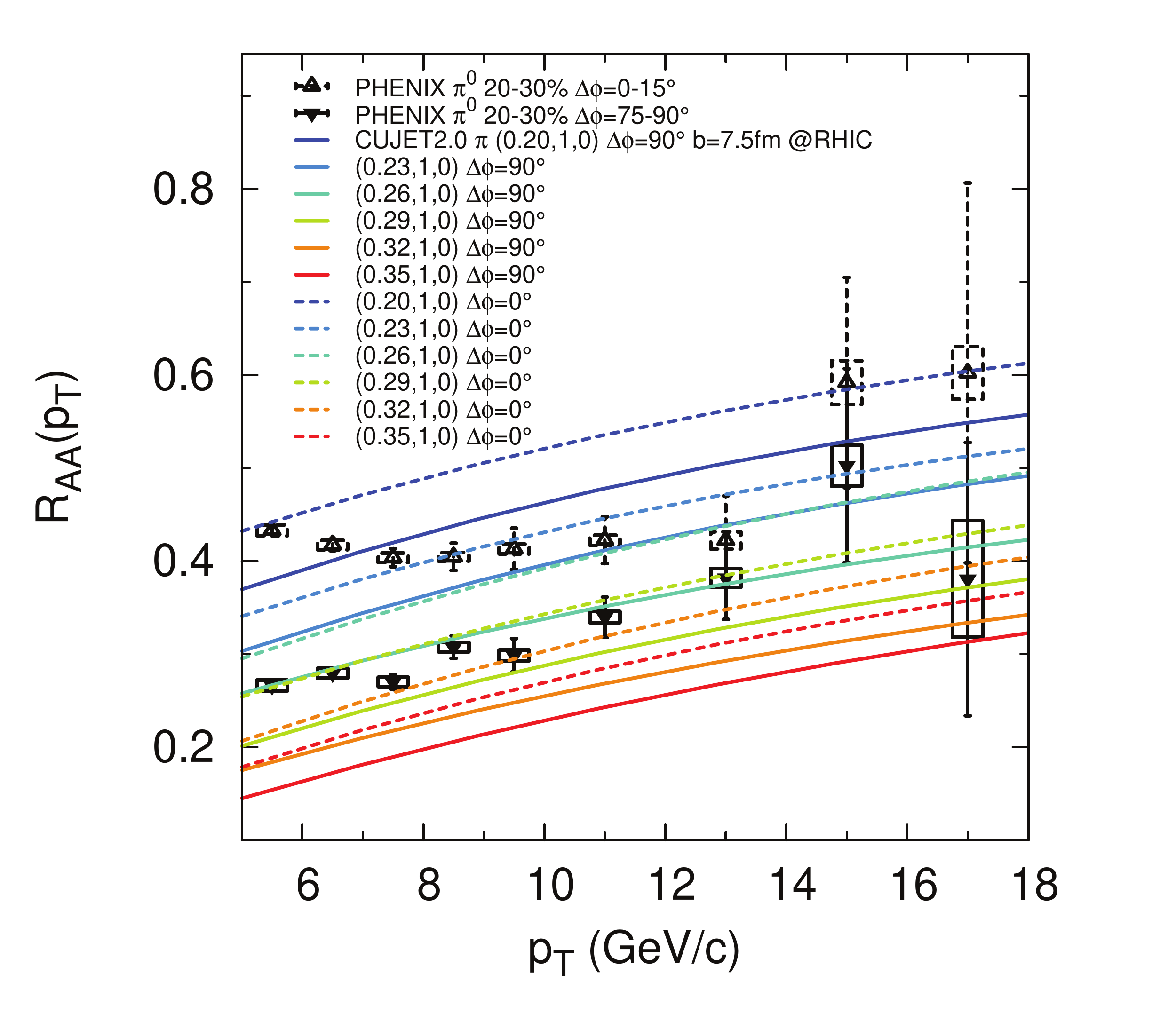}
\includegraphics[width=0.3\textwidth]{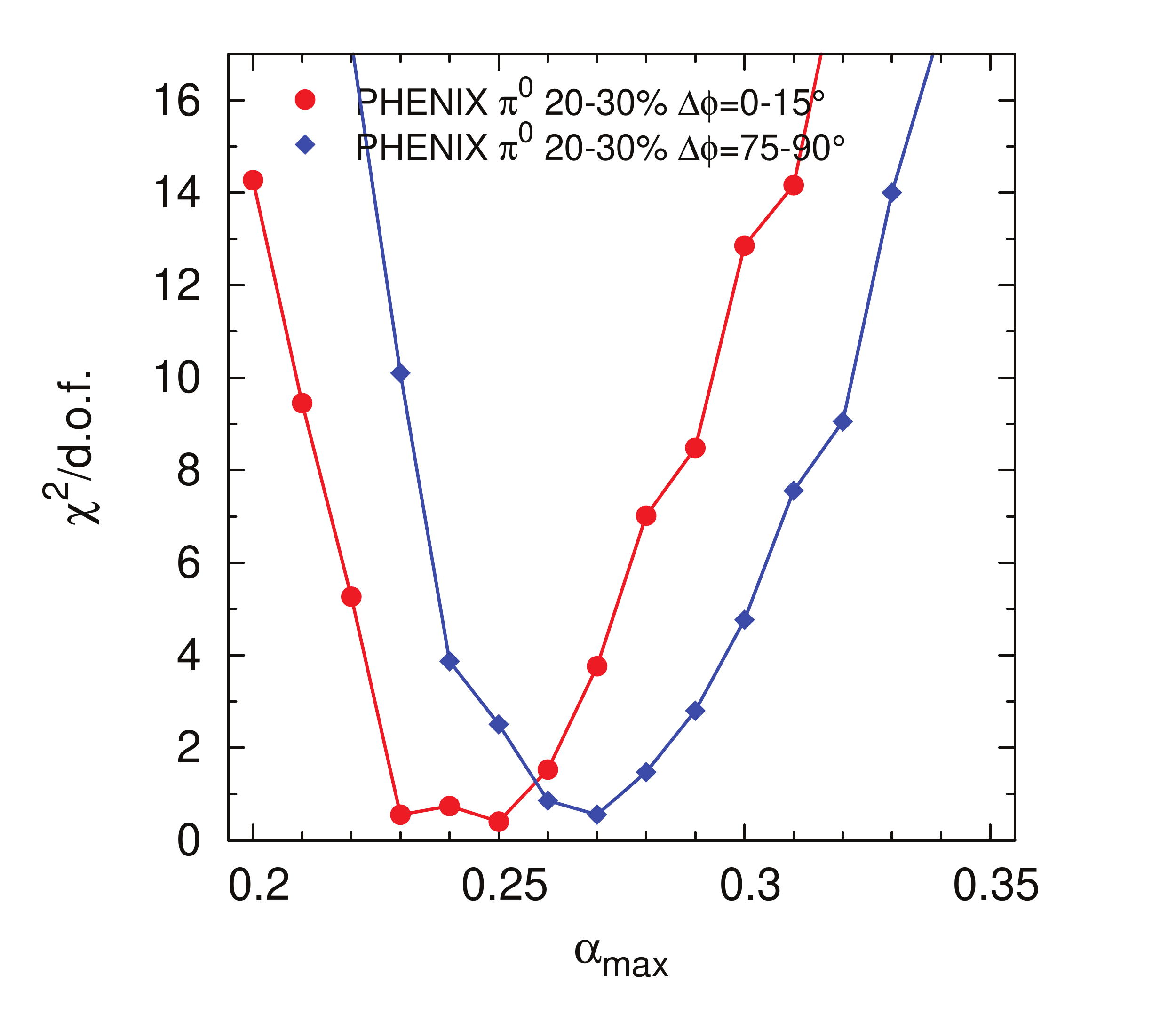}
\includegraphics[width=0.3\textwidth]{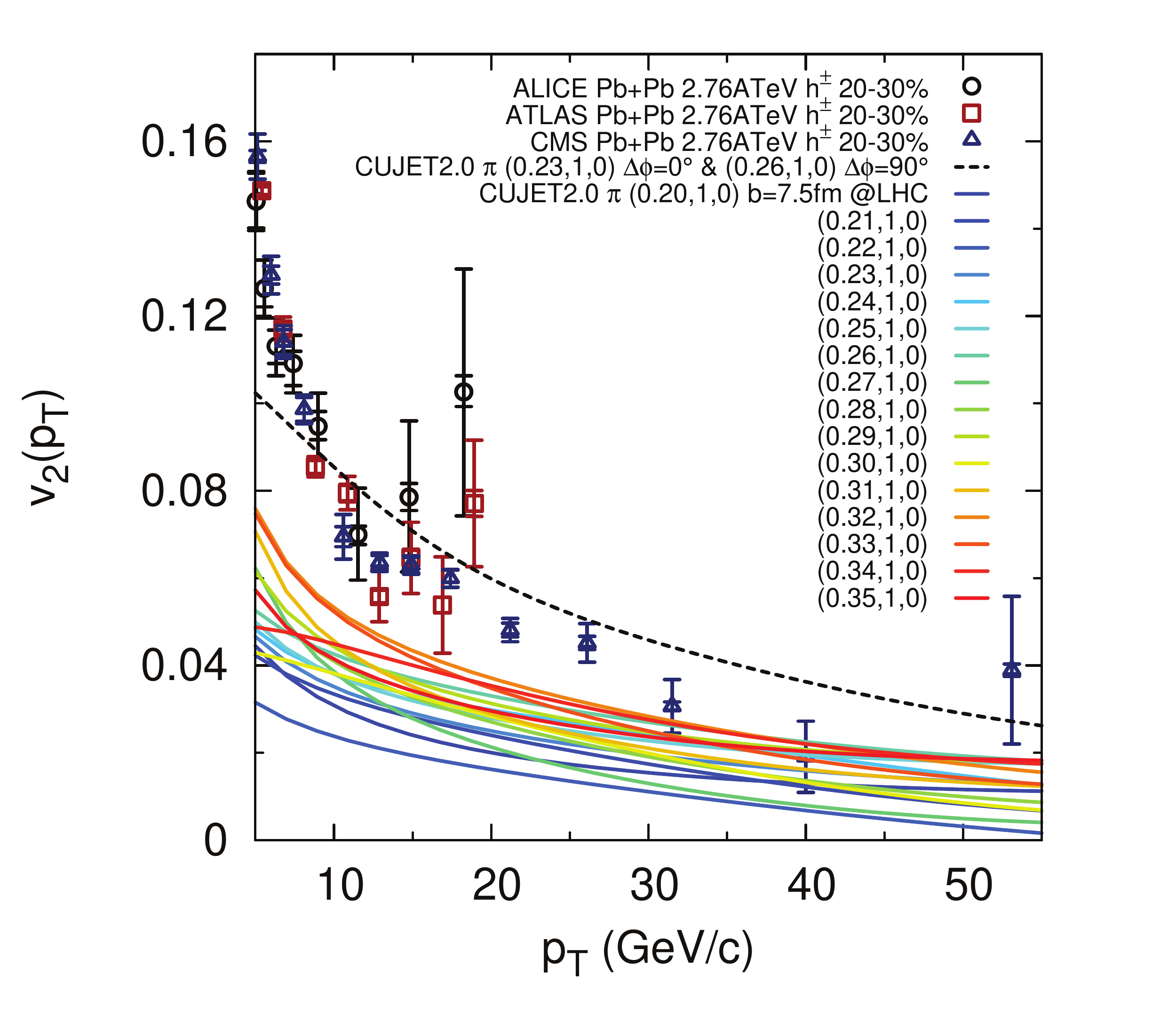}
\caption{
		(Left panels) CUJET2.0 $\pi$ $R_{AA}^{in}$ ($\Delta \phi = 0^{\circ}$, dashed curves) and $R_{AA}^{out}$ ($\Delta \phi = 90^{\circ}$, solid curves) versus $ p_T $ for Au+Au 200 AGeV b=2.4fm (left top), 7.5fm (left bottom) calculated in the dynamical HTL $(f_E,f_M)=(1,0)$ scenario with \amax~varies from 0.20 to 0.35. The bulk evolution profile being used is the same as in Fig.~\ref{fig:Multi_Alf_HTL}. PHENIX \cite{PHENIX} measurements of $\pi^0 R_{AA}$ in Au+Au $\sqrt{s_{NN}}=200 \mbox{GeV}$ with centrality 0-10\% (top left) and 20-30\% (bottom left), and reaction plane $\Delta \phi = 0-15^{\circ}$ (dashed black), $75-90^{\circ}$ (solid black) are compared. (Middle panels) \chisq~vs \amax~for $R^{in/out}_{AA}$ calculated from left panels. $\alpha_{max}=0.26$ $R_{AA}^{in}$ and $\alpha_{max}=0.29$ $R_{AA}^{out}$ for $b=2.4$ fm,  $\alpha_{max}=0.23$ $R_{AA}^{in}$ and $\alpha_{max}=0.26$ $R_{AA}^{out}$ for $b=7.5$ fm generate experimentally compatible $\pi^0$ $R^{in/out}_{AA}(p_T)$ at RHIC, {\it while ensuring fits to average $R^{\pi,h}_{AA}$ in RHIC and LHC at $ \chi^2/d.o.f.<2 $ level}, cf. Fig.~\ref{fig:Multi_Alf_HTL}. (Right panels) Extrapolation to LHC $h^\pm v_2(p_T)$ in Pb+Pb 2.76 ATeV b=2.4fm (top right), 7.5fm (bottom right) collisions after constrained CUJET2.0 with RHIC $R_{AA}^{in/out}(p_T)$. Results with 10\% azimuthal variation for path averaged \amax~(dashed black) are in agreements with ALICE ($v_2\{4\}, |\eta|<0.8$) \cite{ALICE}, ATLAS ($|\eta|<1$) \cite{ATLAS} and CMS ($|\eta|<1$) \cite{CMS} measurements in both central and semi-peripheral AA collisions.
		}
\label{fig:RAA_in_out}
\ec
\end{figure*}

The reaction plane dependent $\pi^0$ quenching pattern, $R^{in/out}_{AA}(p_T)\approx R_{AA}(p_T)[1\pm 2 v_2(p_T)]$, at RHIC is calculated in CUJET2.0 and shown in the left panels of Fig.~\ref{fig:RAA_in_out},
with corresponding $\chi^2/d.o.f.(\alpha_{max})$ demonstrated in the middle panels. Choose $\alpha_{max}=0.26$ for $R_{AA}^{in}$ and $\alpha_{max}=0.29$ for $R_{AA}^{out}$ in $b=2.4$ fm collisions, $\alpha_{max}=0.23$ for $R_{AA}^{in}$ and $\alpha_{max}=0.26$ for $R_{AA}^{out}$ in $b=7.5$ fm collisions, the reaction plane dependent pion suppression pattern computed from the CUJET2.0 model are consistent with RHIC measurements. Since $R_{AA}(p_T)\approx [R^{in}_{AA}(p_T;\alpha^{in}_{max})+R^{out}_{AA}(p_T;\alpha^{out}_{max})]/2\approx R_{AA}(p_T;(\alpha^{in}_{max}+\alpha^{in}_{max})/2)$, with this set of \amax's, {\it average $R^{\pi,h}_{AA}$ has $ \chi^2/d.o.f.<2 $ at both RHIC and LHC} according to the right panels of Fig.~\ref{fig:Multi_Alf_HTL}. Furthermore, extrapolations to LHC elliptic flow $v_2(p_T)$ via inversing $R^{in/out}_{AA}(p_T;v_2(p_T);\alpha^{in/out}_{max})$ are also in considerable agreements with data, as shown in right panels of Fig.~\ref{fig:RAA_in_out}. Therefore, with as less as 10\% azimuthal variation of path averaged \amax, CUJET2.0 results are consistent with both \raa~and \vtwo~data at both RHIC and LHC at the level of average $ \chi^2_{R_{AA},v_2}/d.o.f.<2 $

The physics underneath the azimuthal angle dependence of the path averaged strong coupling strength is two fold: Firstly, in the setup of the running coupling in the DGLV opacity series, the temperature scale only appears in the thermal running of the Debye screening mass. For $\alpha_{max}=0.26$, the minimal running scale $Q_{min}\sim2.9$GeV, it means in this situation the temperature scale has no impact on the running coupling, which is unnatural. If for example a non-perturbative near $T_c$ enhancement of $\alpha_s$ exists, local temperature effects would lead to an amplification of late time energy loss and hence a larger $v_2$, and it would manifest itself in the present CUJET2.0 model as a dependence of path averaged coupling strength on the azimuthal angle. Secondly, if a 10\% change in the in- and out-of plane path averaged coupling strength can induce 100\% increase of $v_2$, then there is a possibility that the originally under-predicted $v_2$ in the model comes from uncertainties in the hydro evolution profile, because high-$p_T$ $v_2$ comes roughly half from the energy loss and half from the hydro bulk, and at current stage only the smooth profile from VISH2+1 is used.

In addition, the $\alpha_{max}$'s azimuthal dependence has a clear pattern: at $\tau_0$, the length of the medium in the $b=2.4$ fm and $b=7.5$ fm collision along the $\phi=0^{\circ}$ and $\phi=90^{\circ}$ direction can be approximately ordered as $7.5{\rm fm} + 0^{\circ} < 7.5{\rm fm} + 90^{\circ} \approx 2.4{\rm fm} + 0^{\circ} < 2.4{\rm fm} + 90^{\circ}$, and the best fit \amax~in corresponding situations is $0.23<0.26=0.26<0.29$ --- a longer path requires a stronger coupling for correctly predicting the high $p_T$ single particle $v_2$.


\section{Summary}
\label{summ}

In the DGLV based CUJET2.0 \cite{CUJET2.0} model, for semi-peripheral A+A collisions, with only $\sim$10\% variation of the average coupling strength for paths in-plane versus out-of-plane, the azimuthal asymmetry of jet quenching can be accounted for at both RHIC and LHC.


\label{acknowledge}

We thank Barbara Betz, Andrej Ficnar, Jinfeng Liao, Chun Shen and Xin-Nian Wang for many useful discussions. Support for this work under U.S. DOE Nuclear Science Grants No.DE-FG02-93ER40764 and No.DE-AC02-05CH11231 and OTKA grant NK106119 is gratefully acknowledged.












\end{document}